\documentstyle[11pt]{article}
\def\p{\partial} \def\D{{\cal D}}  
\def\dpp{\p^{++} }  \def\Dpp{\D^{++} }
\def\e{\epsilon} \def\g{\gamma} \def\n{\nabla} \def\d{\delta}
 \def\s{\sigma}   
\def\t{\vartheta}  \def\bt{{\bar \vartheta}} 
\def\b{\beta} \def\a{\alpha} \def\l{\lambda}  \def\f{\varphi}
\def\G{\Gamma} \def\L{{\cal L}}
   
\def\dd{{\dot \delta}}
\def\ba{{\breve\alpha}} \def\bb{{\breve\beta}} 

\def\bi{{\breve k}} 
\def\bA{{\breve A}} \def\bB{{\breve B}} \def\bC{{\breve C}}
  
\def\o{\omega} 
\def\sd{self-dual } \def\sdy{self-duality } 
\def\sfld{superfield }  \def\hk{hyperk\"ahler }
 
 \def\h{harmonic } \def\ss{superspace }
\def\sg{supergravity } \def\ssdy{super self-duality }
\def\half{{1\over 2}}
\def\der#1{{\partial \over \partial #1}}
\def\dd#1#2{{\partial #1 \over \partial #2}}
\def\be{\begin{equation}}
\def\r#1{(\ref{#1})} \def\la#1{\label{#1}}
\def\ee{\end{equation}}
\def\arr{\begin{array}{rll}}
\def\ea{\end{array}}
\begin{document}
\rightline{\today}
\rightline{hep-th/9501061}
\rightline{Dubna, JINR-E2-94-384}
\vskip 15 pt
\centerline{\large SELF-DUAL SUPERGRAVITIES}
\vskip 15 pt
\centerline{Ch. Devchand  and  V.  Ogievetsky} \vskip 10 pt
\centerline{Joint Institute for Nuclear Research, 141980 Dubna, Russia}
\vskip 15 pt
\begin{quote}
The N-extended supersymmetric self-dual Poincar\'e supergravity
equations provide a natural local description of supermanifolds
possessing hyperk\"ahler structure. These equations admit an economical
formulation in chiral superspace. A reformulation in harmonic superspace
encodes self-dual supervielbeins and superconnections in a graded
skew-symmetric supermatrix superfield prepotential satisfying generalised
Cauchy-Riemann conditions. A recipe is presented for extracting explicit
self-dual supervielbeins and superconnections from such `analytic'
prepotentials. We demonstrate the method by explicitly decoding a
simple example of superfield prepotential, analogous to that corresponding
to the Taub-NUT solution. The superspace we thus construct is an
interesting  $N=2$ supersymmetric deformation of flat space, having flat
`body' and constant curvature `soul'.
\end{quote}
\vskip 15 pt
\section{Introduction}
Recently we presented a version \cite{revisited} of Penrose's `nonlinear
graviton' \cite{penrose} construction in harmonic space language. This
encodes all \sd solutions to Einstein equations in an analytic prepotential
in harmonic space \cite{annals}.  This reformulation of the curved twistor
construction yields a transparent method for the explicit construction
of \sd metrics and connections. In four dimensions the \sdy equations are well
known to be differential equations encoding hyperk\"ahlerity conditions and
the aim of the present work is to generalise our construction to
N-extended Poincar\'e supergravities, yielding \hk {\em superspaces}
having the supergroup $SU(N|2)$ as the holonomy group.
Our method is N-independent, treating all extensions of \sd gravity
on an equal footing.

Let us recall the $N=0$ \sdy constraints which we want to supersymmetrise.
In four dimensional space, the Riemann tensor has spinorial decomposition
$$[{\cal D}_{\b b}, {\cal D}_{\alpha a}] =
\epsilon_{a b} R_{\alpha \b} + \epsilon_{\alpha \b  } R_{a b},$$
with
$$\arr
R_{\alpha \b  } &\equiv &
C_{(\alpha\b  \gamma\delta)} \Gamma^{\gamma\delta}
 + R_{(\alpha\b  )(cd)} \Gamma^{cd} +{1\over 6}R \Gamma_{\alpha\b  } ,\cr
R_{a b} &\equiv &
C_{(abcd)}\Gamma^{cd} + R_{(\gamma\delta)(ab)} \Gamma^{\gamma\delta}
+ {1\over 6} R \Gamma_{ab}, \ea $$
where round brackets denote symmetrisation and, in this spinor notation,
$C_{(abcd)} (C_{(\alpha\b  \gamma\delta)}) $ are the (anti-) self-dual
components of the Weyl tensor, $R_{(\alpha\b  ) (c d)}$  are the components
of the tracefree Ricci tensor, $R$ is the scalar curvature,
$(\Gamma^{\gamma\delta},\Gamma^{cd})$ are generators
of the tangent space gauge algebra.
The usual N=0 \sdy conditions for the Riemann tensor,
may therefore be written in spinorial notation in the form of the
constraints
\begin{equation}    R_{a b}=0\; ,\quad {\mbox i.e.}
\quad  [\D_{\b b}, \D_{\a a}] = \e_{a b} R_{\a \b} =
C_{(\a\b\g\d)} \G^{\g\d}\;  ,\label{1}\ee
where $\a,\b$ are indices of a {\em local} tangent-space $su(2)$ algebra
(generators $\G^{\g\d}$),
$a,b$ are indices of a {\em rigid} tangent-space $su(2)$ algebra and the
covariant derivatives $\D_{\a a}$ take values in the former (local) $su(2)$
algebra. With one of the simple parts of the rotation group thus `de-gauged',
the \sdy of the Riemann tensor is automatic  and \r1 reduces to
conditions for torsion-free $\D_{\a a}$. (Conversely, \sdy, being equivalent
to the vanishing of all curvature coefficients of the generators
$\Gamma^{cd}$, allows this `self-dual gauge'). The curvature and connection
thus taking values in an $su(2)$ algebra, the manifold is manifestly a
\hk one with holonomy group $SU(2)$.

To supersymmetrise \r1 we consider supercovariant derivatives on
N-extended Poincar\'e superspace  $\{ \D_{\a a}, \D_{i a}, \D_\a^i\; \}$
and generalise \r1 to {\it superfield} relations implying {\it superspace}
constraints on the vielbeins and connections of $\D_{\a a}$ (which now
depend on the coordinates of superspace
$\{ (x^{\mu a}, \bt^{m a}), \t^\a_m \}$).
We maintain the tangent-space subalgebra labeled by the spinor
indices $a,b,...$ as a {\em rigid} symmetry algebra and
$ m=1,..N$ is the index of the internal automorphism group of the
N-extended Poincar\'e supersymmetry algebra
\setcounter{equation}{0}
\renewcommand\theequation{2\alph{equation}}
\begin{equation}     \{\D^i_\a, \D_{j a} \} = 2\d^i_j  \D_{\a a} .\ee
In N-extended superspace, the constraints \r1 need to be augmented
by further constraints amongst the supercovariant derivatives.
The simplest set of constraints generalising \r1 consistent with
(2a) is the following set
\begin{eqnarray}
\{\D^j_\b, \D^i_\a  \} =& 0                \\[0pt]
[\D^i_\b,  \D_{\a a} ]  =& 0                \\[0pt]
\{\D_{j b}, \D_{i a}\} =&  \e_{a b} Z_{i j}  \\[0pt]
[\D_{i b}, \D_{\a a}] =&  \e_{a b} W_{\a i}   \\[0pt]
[\D_{\b b}, \D_{\a a}] =& \e_{a b} R_{\a \b}, \end{eqnarray}
\setcounter{equation}{2}
\renewcommand\theequation{\arabic{equation}} \noindent
where $R_{\a\b}$ is the standard dimension $2$ \sd Riemann curvature
in extended superspace and $Z_{ij}, W_{\a i}$ are respectively the
dimension $1$
and dimension ${3\over 2}$ components of the supercurvature.
Note that we consider only terms proportional to the antisymmetric
invariant $\e_{ab}$ on the right of (2), taking precisely the vanishing
of all terms not proportional to $\e_{ab}$, together with the condition
that the superconnections and supercurvatures in (2) do not take values in
the rigid $su(2)$ subalgebra of the tangent-space superalgebra,
to be the supersymmetric generalisation of the \sdy conditions \r1.
The constraints (2) are the simplest generalisation of \r1 consistent
with the $N-$extended Poincar\'e supersymmetry algebra, containing the
minimal number of superfields needed to describe the \sd graviton
supermultiplet.  They are identical to the constraints obtained in the
`ungauged' limit of those given in \cite{siegel}.
Note that we can treat any possible extended supersymmetry on
the same footing; there is no N-dependence in the form of these equations.
This is a characteristic feature of self-dual (gauge and gravity) theories,
which distinguishes them from their non-self-dual counterparts.
As for \sd supersymmetric gauge theories \cite{sym} our considerations
are good for complexified superspace or for real superspaces with `bodies'
of signature (4,0) or (2,2) (with appropriate handling of the latter as
a restriction of complexified superspace). For concreteness however, we
shall deal with the real Euclidean version.

In the next section we show that the above \ssdy constraints take on a more
economical form in chiral superspace, which is amenable to reformulation in
harmonic superspace (section 3). The latter superspace has a special
class of local coordinate frames called analytic frames, in which the local
diffeomorphism group preserves an analytic subspace. In section 4 we write down
the equations for the analytic frame supervielbeins and superconnections.
The advantage of the abovementioned reformulation in harmonic superspace
is that the non-zero supercurvature components get rearranged so as to single
out some commuting subset of covariant derivatives. Frobenius' theorem may
then be used to `flatten' the latter subset of covariant derivatives.
In section 5  we discuss the corresponding `Frobenius' gauge, in which the
equations become manifestly soluble in terms of an analytic supermatix of
unconstrained prepotentials (section 6). In section 7 we give a recipe for the
extraction of the geometrical data (supervielbeins and superconnections) from
the latter harmonic space data and we conclude with a simple explicit example
(section 8) of our construction.

\section{Super \sdy in chiral superspace}

The constraints (2) admit an economical reformulation in {\em chiral}
superspace. Indeed, super-Jacobi identities involving $\D^j_\a$ imply the
important condition that the Riemann tensor superfield, $R_{\a\b}$, is
{\em chiral}, \begin{equation}     \D^j_\g  R_{\a\b} =0, \ee and further, that
the superfield curvatures $Z_{ij}, W_{\a i} $ and $ R_{\a\b}$ are
related to each other thus:
\begin{equation}    R_{\a \b} = {1\over 2N} \D^j_\b W_{\a j} ,
\quad \D^j_\b W_{\a i} = {1\over N} \d^j_i \D^k_\b W_{\a k},
\quad (N-1) W_{\a i} = \D^j_\a Z_{i j}.\label{bianchi}\ee
This redundancy of the description (2) manifests itself in a
local spinorial (i.e. odd) tangent-space symmetry:
\begin{equation}    \d \D_{ia}=\eta^\a_i \D_{\a a} ,\quad  \d \D^i_\a =0,
\quad \d \D_{\a a}=0 \label{eta}\ee which preserves the forms of (2a-c) and
provides the transformation laws
\begin{equation}    \d Z_{i j} = \eta^\a_{[j} W_{\a i]}  ,\quad
  \d W_{\a i} =\eta^\b_i R_{\a \b}      ,\quad
  \d R_{\a \b} =0 .\label{trans}\ee
This symmetry is a novelty for the \ssdy equations (2); it does not
exist in the non-\sd theory; and allows a very economical reformulation
of the constraints (2) in chiral superspace.
Indeed, comparing the above transformations with the
algebra (2) and the relationships \r{bianchi},
it is clear that the covariant derivative $\D^j_\a$ is actually the
generator of these transformations. In other words, the parameter
$\eta^\a_i$ is just a parameter of $\t$--translations:
$\d \t^\a_i = \eta^\a_i$.
Now, the form of the first three constraints (2a-c) actually allows
us to use this additional odd invariance parametrised by $\eta^\a_i$
in order to gauge-away all $\t^\a_i$--dependences. Namely, (2b) implies
the existence (by Frobenius'theorem) of a chiral basis in which (2a-c)
have solution
\begin{equation}   \begin{array}{lcl}
	    \D^i_\a    &=&  \der{\t_i^\a }                  \cr
            \D_{i a}   &=&  \n_{i a}  +  2 \t_i^\a \n_{\a a}  \cr
	    \D_{\a a}  &=&  \n_{\a a}              \label{de-chiral} \ea\ee
where  $\n_{A a} = \{ \n_{i a},\n_{\a a} \}$ are supercovariant derivatives
in {\it chiral} superspace with coordinates
$z^{M a}= \{ x^{\mu a}, \bt^{m a} \}$. In other words
they may be expressed in terms of {\em chiral} supervielbeins and
superconnections thus:
\begin{equation}    \n_{A a} = E^{M b}_{A a}(z) \der{z^{M b}} + \o_{A a}(z)\ .
\label{bein}\ee
The inverse supervielbein defines the differential one-forms
\begin{equation}    E^{A a} = E^{A a}_{M b} dz^{M b}\  .\ee

Inserting \r{de-chiral} in (2d-f) yields the following equivalent
system of constraints in {\it chiral superspace} :
\setcounter{equation}{0}
\renewcommand\theequation{11\alph{equation}}
\begin{eqnarray}
\{\n_{j b}, \n_{i a}\} =& \e_{a b}  R_{i j} \\[0pt]
[\n_{i b}, \n_{\a a}] =& \e_{a b}  R_{\a i}   \\[0pt]
[\n_{\b b}, \n_{\a a}] =& \e_{a b} R_{\a \b},
\end{eqnarray}
where the supercurvature components $R_{i j}, R_{\a j}, R_{\a \b}$
are chiral (i.e. independent of $\t$) superfields; and
we have made the identifications
\setcounter{equation}{11}
\renewcommand\theequation{\arabic{equation}}
\begin{equation}    W_{\a i}= R_{\a i} + 2 \t_i^\b R_{\a \b} ,\qquad
Z_{i j} = R_{ij} + 2 \t_{[i}^\a R_{\a j]} + 4 \t_i^\a\t_j^\b R_{\a\b}
\label{nc}\ee
which solve \r{bianchi}.

Using a superindex $A=(\a, i)$ of the superalgebra
$su(N|2)$ having an $su(N)\times su(2)$ even part, where
$\a$ is an $SU(2)$ spinor index and $i$ is an $SU(N)$ vector
index, eq.(3) takes the manifestly supersymmetric compact form
\begin{equation}    [\n_{B b}, \n_{A a}\} = \e_{a b} R_{AB}\;  .\label{ssd}\ee
This form displays manifest $su(2)\times su(N|2)$ tangent space covariance,
of which we maintain the $su(2)$ factor as a rigid symmetry group, gauging
the local $su(N|2)$ symmetry. So the Riemann supercurvature has components
$R_{AB} = R_{ABC}^{~~~~~D} \G_D^C $, where $\G_D^C$ are the generators of
the tangent space superalgebra $su(N|2)$. Since both superconnection and
supercurvature take values in this factor of the tangent space algebra,
the holonomy superalgebra is manifestly $su(N|2)$.

In fact the constraints written in this manner reminiscent of $N=0$ \sdy may
immediately be generalised to equations describing $d>4$ \hk superspaces.
Generalising the indices $A,B$ to superindices of  $osp(N|2m)$ yields a
supersymmetrisation of the type of higher dimensional \sdy conditions
considered in \cite{ward}. Together with the proviso that the supercurvatures
and superconnections take values in this  local $osp(N|2m)$ algebra (i.e. the
remaining $su(2)$ tangent space symmetry remains rigid), these equations
provide a local description of  $(4m|2N)$-dimensional \hk superspaces
(allowing N independent supersymmetries). The $m=1$ case of $osp(N|2)$
holonomy corresponds to a reduction (symmetry-breaking) of the
presently discussed $su(N|2)$  theory.

The rigidity of the $SU(2)$ factor of the tangent space group is
particularly important. It yields the extra freedom allowing the construction
of invariants transforming non-trivially under this $SU(2)$ (since only the
superindices of the local supergroup need to be summed over).
In particular, the three independent complex structures may be
constructed from the supervielbein thus:
\begin{equation}
  (I^k )^{M a}_{N b} = -i E^{A c}_{N b} (\s^k )^d_c  E^{M a}_{A d}\
\label{cs}.\ee
They may easily be seen to satisfy the algebra of the imaginary quaternion
units, $ I^k I^m = -\d^{km} + \e^{kmn}I^n$, and are covariantly constant
in virtue of the zero-supertorsion conditions \r{ssd}.

Note that in chiral superspace the \ssdy equations have the same
super-covariant form \r{ssd}, irrespective of whether we start from the the
constraints (2) or whether we include additional gauge fields in the
non-chiral formulation \cite{siegel}.  The difference is entirely absorbed in
the
transformation to the chiral basis.  Although \r{ssd} is a less redundant
description of \ssdy than (2), the superfield components of $R_{AB}$ (i.e. $
R_{i j}, R_{\a j}, R_{\a \b}$) are still not independent of each other, since
they satisfy the super Jacobi identities \begin{equation}
  \n_{[Ca} R_{A)B} = 0 .\ee The
independent supercurvature components, i.e. the solutions of these
super-Jacobi identities, are the \sfld curvatures of the \sd \sg multiplet
\cite{siegel}.  (We denote graded skew-symmetrisation of superindices by
$[~~)$, i.e. $T_{[AB)} = T_{AB} - (-1)^{AB} T_{BA}$, where we use the
notational convention that each letter appearing in the exponent of $(-1)$
assumes the value 0 or 1 according to whether the corresponding index is even
or odd).

In this chiral superspace, the $N$-extended diffeomorphism supergroup
is realised by the {\it chiral} diffeomorphisms
\begin{equation}   \begin{array}{rrrl} \d z^{M a} = \tau^{M a}(z) ,\quad
{\mbox i.e.}&\d x^{\mu a} &=& \tau^{\mu a}(x, \bt),\cr
            &\d \bt^{i a} &=& \tau^{i a}(x, \bt),\label{diff}\ea\ee
where we denote the coordinate of chiral superspace
using the superworld index $M=(\mu, m)$ by  $ z^{M a} =
\{ (x^{\mu a}, \bt^{m a}) \}$. Since half the rotation group is rigid,
the coordinates $x, \bt$ have only one world-spinor index, the index $a$ being
identified with the corresponding tangent-space index. Similarly, local
$su(N|2)$ tangent transformations also have chiral parameters,
$\tau^B_A(z)$. The transformation of the superconnection components
\begin{equation}     (\o_{A a} )_B^C \rightarrow
 (-1)^{D(B+E)} \tau^D_A  \tau^E_B  (\o_{D a})_E^F (\tau^{-1} )^C_F
 -  \tau^D_A E_{D a}^{M m}\p_{M m} \tau^E_B  (\tau^{-1} )^C_E       \ee
provides covariant transformation rules for $\n_{A a}$ \r{bein} as well
as for the supercurvature and supertorsion components (c.f. \cite{deWitt})
\begin{equation}   \arr  \e_{ab} R_{ABC}^{~~~~D} =
E_{B b}^{M m}\p_{M m} (\o_{A a} )_C^{~D}
	 &+&    (-1 )^{A(F+C)} (\o_{B b} )_C^{~F} (\o_{A a} )_F^{~D} \\[7pt]
	 &+&    (\o_{B b} )_A^{~F} (\o_{F a} )_C^{~D}
        - (-1 )^{AB} (Aa \leftrightarrow Bb)  \ea\label{curv}\ee
$$ T^{~~~~~Cc}_{Bb,Aa} \n_{C c} = E_{B b}^{M m}\p_{M m} E_{A a}^{N n}\p_{N n}
	 +  (\o_{B b} )_A^{~C} E^{M m}_{C a} \p_{M m}
        - (-1 )^{AB} (Aa \leftrightarrow Bb)\	 .$$
The sign factors in the above definition of the supercurvature are crucial
in the proof of covariance.

\section{Super \sdy in \h \ss}\label{har}

The system \r{ssd} has precisely the same form as the $N=0$ \sdy conditions
except that the indices $\a,\b$ of the latter have become superindices
$A,B$. We may therefore reformulate these equations in harmonic \ss
by closely following the treatment of \cite{revisited}. In suitable local
coordinates the system \r{ssd} becomes manifestly soluble. We shall
describe the main steps of our procedure, referring to \cite{revisited}
for details and proofs. The rigid SU(2) tangent-space symmetry allows
us to reformulate \r{ssd} in \h \ss. Consider  $S^2 = {SU(2)\over U(1)} $
harmonics \cite{har}
$\{ u^{\pm a}~ ;~ u^{+a}u^-_a =1,~ u^\pm_a \sim e^{\pm\gamma} u^\pm_a \}$,
where a is the spinor index of the rigid $SU(2)$ and $\pm$ denote U(1) charges.
These harmonics allow us to define a special coordinate system in harmonic
superspace, the {\em central coordinates},
$\{ z^{M\pm} = z^{Ma} u^\pm_a , u^\pm_a \}$, which are linear in the
harmonics. Using these we define \h \ss covariant derivatives in
the {\em central frame} thus:
\begin{equation}    \D^\pm_A \equiv D^\pm_A + \o^\pm_A = u^{\pm a} \n_{A a},
\quad \D^{++}= \p^{++} = u^+_a\der{u^-_a}\   .\label{c}\ee
\setcounter{equation}{0}
\renewcommand\theequation{20\alph{equation}}

{\it The following system in \h \ss is equivalent to the \ssdy constraints
\r{ssd}:}
\begin{eqnarray}
 [\D^+_A, \D^+_B \} &=&0 \\[0cm]
 [\D^{++}, \D^+_A] &=& 0 \\[0cm]
 [\D^+_A, \D^-_B \} &=& 0 \quad\hbox{(modulo  $R_{AB}$)} \\[0cm]
 [\D^{++}, \D^-_A ] &=&\D^+_A,
\end{eqnarray}
The proof of equivalence in this
central basis follows by linear algebra from the requirement that all
the supertorsion constraints implicit in \r{ssd} be implied by (20).
The central frame \r c, with its characteristic
feature that the harmonic derivative $\Dpp $ is a partial derivative
acting only on harmonics and is connection-less, whereas the
derivative $\D^\pm_A$, in virtue of (20a), has a pure-gauge form of
connection,
\setcounter{equation}{20}
\renewcommand\theequation{\arabic{equation}}
\begin{equation}    \D^\pm_A = D^\pm_A -  D^\pm_A \f\f^{-1}  \; ,\la{conn}\ee
has the privilege of a manifest equivalence, $\r{ssd} \Leftrightarrow (20)$.
The system (20), however, is covariant under diffeomorphisms,
allowing a choice of any other local coordinate system
$\{z^{M\pm}=z^{M\pm}(z^{Ma} u^\pm_a , u^\pm_a) , u^\pm_a \}$.
It is also covariant under local $su(N|2)$ tangent frame (i.e. supergauge)
transformations under which  $\f$ in \r{conn} transforms thus:
\begin{equation}
 \f^{\bA}_A \rightarrow \tau_A^B(z^{Ma}) \f_B^\bA ,\label{tau}\ee
with local parameters $\tau^B_A(z^{Ma})$ of the conventional central frame
tangent supergroup.

The advantage of reformulating \ssdy in the form (20) is that the \sd
supercurvature components have now been rearranged so as to make two
dimensional
flat subspaces manifest, allowing us to use Frobenius' theorem in order to
make the transformation to another special coordinate system, the
{\em Frobenius coordinates} in which the explicit solubility of (20)
actually becomes manifest. Our strategy is to solve (20) in the latter
coordinates of manifest solubility and then to perform a coordinate
transformation back to the central coordinates of manifest
$\r{ssd} \Leftrightarrow (20)$ equivalence in order to extract the solution
to the original system (2) from the solution of (20). The central basis
therefore acts as a bridge between the original superspace and \h \ss.

The abovementioned `Frobenius' coordinate system belongs to a very
useful class of local coordinates for \h \ss, namely
{\it analytic coordinates} or h-coordinates.
These are coordinate systems $\{ z^{M\pm}_h, u^\pm_a \}$ in which the
group of diffeomorphisms preserves an `analytic' subspace in the sense of
the following transformation rules
\begin{equation}    \d z^{M+}_h = \lambda^{M+}(z^+_h, u)\  ,
\label{a_diff+}\ee while
\begin{equation}
 \d z^{M-}_h = \lambda^{M-}(z^+_h, z^-_h, u)\   ,\label{a_diff-}\ee
which clearly leave the `analytic' subspace with coordinates
${z_h^{M+}, u^\pm_a}$ invariant. The coordinate system
$\{ z^{M\pm}_h, u^\pm_a \}$ defines a basis of derivatives covariant under
these transformations. In particular,
\begin{equation}
 \D^+_A \rightarrow \D^+_\bA = (\f^{-1})^A_\bA (D^+_A z^{M-}_h) \p^+_{hM}
\equiv f^M_\bA \der{z^{M-}_h}\  ,\label{an2}\ee
where $f^M_\bA$ is a supervielbein. In other words,
the h-coordinates as functions of the central ones need
to satisfy the relations \begin{equation}
 \D^+_A z^{M+}_h = 0\  ,\label{hol}\ee
which prevent the appearance of $\p^-_{hM} \equiv \der{z^{M+}_h}$
on the right of \r{an2}.
Then the conditions $\D^+_\bA \Psi = 0$  imply
(for invertible vielbein $f^M_\bA$) the `analyticity' of $\Psi$, i.e.
$\p^+_{hM}\Psi = 0.$  This independence of $z_h^{M-}$ is what we mean
by `analytic', irrespective of whether our coordinates are taken to be
real or complex.
(We take our superspace coordinates $z^{M a}, z_h^{M\pm}$
to be real. The appropriate conjugation  for the
harmonics is discussed in the harmonic space literature, e.g. \cite{har})

Further, in order to make this analyticity concept covariant under
tangent-space transformations as well, in \r{an2} we `gauge away'
the connection in $\D^+_A$ thus:
\begin{equation}    \D^+_A \rightarrow  \f^{-1} \D^+_A \f = D^+_A \la{gt}\ee
and also perform the tangent frame rotation:
$$  D^+_A  \rightarrow D^+_\bA  = (\f^{-1})^A_\bA D^+_A .$$
The analyticity condition $\D^+_\bA \Psi = 0$ is then manifestly
covariant under the local {\em analytic}  transformations
\begin{equation}    \f^{\bB}_B \rightarrow
 \f_B^\bA \lambda_\bA^\bB(z_h^{+M}, u)\  ,\label{19}\ee
with local parameters
$\lambda^\bA_\bB(z_h^{+M}, u)$ of the
analytic frame tangent supergroup.
The matrix $\f$  therefore transforms under local supergauge transformations
according to two distinct realisations of $su(N|2)$ (\r{tau} and \r{19});
so $\f$ clearly plays the role of a `bridge' transforming $\tau$-group
tangent-space superindices ($A,B$) into $\lambda$-group ones ($\bA,\bB $),
distinguished by `breved' superindices.
In the analytic frame covariant quantities are those having only the latter
type of tangent-space indices; and we shall only use such quantities, using
as many $\f$'s as are required in order to convert $\tau$-transforming
indices into $\lambda$-transforming ones; as for $\D^+_A$ above.

The negatively charged covariant derivatives $\D^-_A$, consistently with
\r{a_diff-}, contain derivatives with respect to all the new coordinates:
\begin{equation}
 \D^-_\bA = - e^M_\bA \p^-_{h M} + e^{--M}_\bA \p^+_{h M} + \o^-_\bA ,
\label{9}\ee
where  $e^M_\bA$ is another neutral supervielbein with components
\begin{equation}    e^M_\bA = - (\f^{-1})^A_\bA  D^-_A z^{M+}_h, \label{10}\ee
(the minus sign is chosen so as to have $e^M_\bA=\d^M_\bA$ in
the flat limit) and the doubly-negatively charged supervielbein is
defined by
\begin{equation}
 e^{--M}_\bA = (\f^{-1})^A_\bA  D^-_A z^{M-}_h .\label{11}\ee

The harmonic derivative, in the central frame a partial derivative,
$\D^{++} = \p^{++}$, acquires both vielbeins
and connections in the analytic frame:
\begin{equation}
 \p^{++} \rightarrow \D^{++} = \Delta^{++} + \f^{-1} [\Delta^{++}] \f =
\Delta^{++} + \o^{++}  ,\label{12}\ee  where the
connection is realised as an $su(N|2)$ matrix with `breved' indices,
\begin{equation}
 (\o^{++} )^\bB_\bA = (\f^{-1})^\bB_B \Delta^{++} \f^B_\bA ,\label{18}\ee
and $$ \Delta^{++}
=\p^{++} + H^{++ M+}\p^-_{h M} + (z^{M+}_h + H^{++ M-}) \p^+_{hM} \; .$$
The vielbeins
\begin{equation}     H^{++M+} = \p^{++} z^{M+}_h\;            ,\label{13}\ee
\begin{equation}     H^{++ M-} =\p^{++} z^{M-}_h - z^{M+}_h\; ,\label{14}\ee
are chosen so as to have $H^{++M+} = H^{++M-} = 0$ in the flat limit.

{}From the covariance of the covariant derivatives $\D^\pm_\bA , \Dpp$
under the transformations (\ref{a_diff+},\ref{a_diff-}), we obtain the
following transformation rules for the \h \ss supervielbeins
 \begin{equation}     \d f^M_\bA = f^N_\bA \p^+_{h N}\lambda^{M-} +
  \lambda^\bB_\bA f^M_\bB ,\label{23}\ee
 \begin{equation}     \d e^M_\bA = e^N_\bA \p^-_{h N} \lambda^{M+}
+  \lambda^\bB_\bA e^M_\bB ,\label{24}\ee
\begin{equation}     \d e^{--M}_\bA = - e^N_\bA \p^-_{h N}\lambda^{M-}
+ e^{--N}_\bA \p^+_{h N}\lambda^{M-}
+ \lambda^\bB_\bA e^{--M}_\bB ,\label{25}\ee
\begin{equation}     \d H^{++M+} = \Delta^{++} \lambda^{M+},\label{26}\ee
\begin{equation}     \d H^{++ M-} = \Delta^{++} \lambda^{M-} - \lambda^{M+}\  .
\label{27}\ee

\section{The \ssdy equations for analytic frame superfields}\label{cr}

We now examine the system (20) in an analytic frame,
with covariant derivatives $\D^\pm_\bA , \Dpp$ taking the forms
$$\begin{array}{rll}  \D^+_\bA &=& f^M_\bA \p^+_M \\[7pt]
\D^-_\bA &=& -e^M_\bA \p^-_M+ e^{--M}_\bA \p^+_M + \o^-_\bA \\[7pt]
\Dpp &=& \p^{++} + H^{++ M+}\p^-_{h M}
           + (z^{M+}_h + H^{++ M-}) \p^+_{hM}  + \o^{++}\   .\ea$$
Not all the  supervielbein and superconnection fields in these covariant
derivatives above are independent `dynamical' degrees of freedom. We shall
solve for the superfields in $\Dpp$, which, in a special coordinate gauge,
are the only ones required for the determination of the metric. The remaining
equations are redundant, since the remaining superfields describe the
same degrees of freedom as those in $\Dpp$.

For the superzweibein $f^M_{\bB}$  we have from (20a) the equations
\begin{equation}     f^N_{[\bA} \p^+_{h N} f^M_{\bB)} = 0. \label{28}\ee

The vanishing of the supertorsion coefficients of  $\p^-_{hM}$ in (20c)
and (20b), respectively, requires the vielbeins $e^M_\bA$ and $H^{++ M+}$,
respectively, to be {\it analytic}:
\begin{equation}     D^+_\bA e^M_\bB = 0,\label{29}\ee
\begin{equation}     D^+_\bA H^{++ M+} = 0,\label{30}\ee
The vanishing of the supercurvature in (20b) yields a further analyticity
condition; for the connection $\o^{++}$,
\begin{equation}     D^+_\bA \o^{++} = 0.\label{31}\ee
The solution of this equation is however not independent of the solution of
the previous two analyticity conditions; the equation
\begin{equation}     -\D^{++} e^M_\bA - \D^-_\bA H^{++M+} = 0 ,\label{32}\ee
which is a consequence of the vanishing of the supertorsion coefficients of
$\p^-_{h M}$ in (20d), provides an important constraint amongst the three
analytic superfields $e^M_\bB , H^{++M+}$ and $ \o^{++}$.
Further, these superfields determine  $H^{++M-}$ in virtue of the equation
\begin{equation}      \D^{++} f^M_\bA = D^+_\bA H^{++ M-} \   ,\label{33}\ee
which arises from the requirement of the vanishing of the supertorsion
coefficients of $\p^+_{hM}$  in constraint (20b).

In order to solve (20) it suffices to solve the set of
equations (\ref{28}-\ref{33}). The remaining equations from (20) are
conditions determining consistent expressions for the superfields
$e^{--M}_\bB$ and $\o^-_\bA$, which represent equivalent degrees of
freedom and are therefore redundant (see \cite{revisited}).
The superfield $e^{--M}_\bA$ is determined by the equation following from
the equality of the coefficients of $\p^+_{h M}$ in (20d), namely,
\begin{equation}
 \D^{++} e^{--M}_\bA=- f^M_\bA + \D^-_\bA (H^{++ M-} + z^{M+}_h)
              .\label{34}\ee
The vanishing of supertorsion coefficients of $\p^+_{h M}$ in (20c) yields
\begin{equation}     D^+_\bA e^{--M}_\bB  = \D^-_\bB f^M_\bA \  ,\label{35}\ee
which together with the condition obtained from the requirement that
the antisymmetric part of the supercurvature in (20c) vanishes, i.e.
 \begin{equation}     D^+_{[\bA} \o^-_{\bB)} = 0,\label{36}\ee
determine $\o^-_\bB$, which satisfies the final equation contained in (20),
viz. the vanishing of supercurvature in (20d),
\begin{equation}    \D^{++} \o^-_\bA - \D^-_\bA \o^{++}  = 0 ,\label{37}\ee
automatically, in virtue of (\ref{34}).

\section{The `Frobenius' gauge}\label{gauge}

The set of fields satisfying the system of equations listed in the
previous section possesses the large
class of gauge invariances (\r{23}-\r{27}). We are therefore free to
choose local coordinates partly fixing these gauge degrees of freedom.
In a particularly remarkable coordinate system, eqs.(\ref{28}-\ref{33})
actually becomes manifestly soluble.

Consider the supertorsion constraint (20a), which essentially says, by
Frobenius'
integrability theorem, that $\D^+_\bA$ is gauge-equivalent to the
partial derivative $\p^+_\bA$.

Now, since  $e^M_\bA$ is analytic (\ref{29}), the gauge invariance
(\ref{24}) with analytic parameter $\l^{\bA}_\bB$, allows us to choose
coordinates
$z^{M+}_h$ such that $e^M_\bA$ is also an identity matrix. We therefore
have a coordinate gauge in which both
superzweibeins $f^M_\bA, e^M_\bA$ are identity matrices:
\begin{equation}
 f^M_\bA = \d^M_\bA, \quad e^M_\bA = \d^M_\bA\    .\label{h-flat}\ee
In this special `Frobenius' gauge the distinction between world and
tangent indices has evidently been eliminated and only the set of supervielbein
and connection fields
$\{ H^{++M\pm}, e^{--M}_\bA,$ $ \o^{++} ,\o^-_\bA \}$ remain,
of which  those contained in $\D^{++}$, namely $\{ H^{++M\pm}, \o^{++} \}$
contain all the dynamical (geometrical) information.

In this gauge, residual gauge transformations have
parameters constrained by relations from (\ref{23},\ref{24}), viz.
$$ \p^-_{h \bA}\l^{M+} + \l^M_\bA  = 0, \quad
\p^+_{h \bA}\l^{M-} + \l^M_\bA  = 0 .$$
So the residual diffeomorphism parameters $\l^{M\pm}(z^+, u)$
are no longer arbitrary but are constrained by the relations
$$ \p^-_{h M}\l^{M+} = 0, \quad \p^+_{h M}\l^{M-} = 0\  ,$$
since the tangent parameters $\l^M_\bA$ are supertraceless. It follows that
the thus constrained $ \l^{M+}$ can be expressed in terms of an unconstrained
doubly charged {\it analytic} parameter:
\begin{equation}
 \l^{M+}_{res}(z^+, u) = (-1 )^{N} \p_{hN}^- \l^{[MN)++}(z^+, u)\
\label{39a}.\ee
These diffeomorphism parameters in turn determine the Lorentz ones, the
residual tangent transformations actually being induced by the world ones:
\begin{equation}
 (\l^M_\bA)_{res} = -\p^-_{h \bA} \l^{M+}_{res}(z^+, u)\label{39b}\ee
As for the remaining $\l^{M-}$ transformations, these have parameters:
\begin{equation}     (\lambda^{M-} )_{res} =
\p^-_{h N} \lambda^{M+}(z^+, u) z^{N -} + \tilde{\lambda}^{M-}(z^+, u)
\label{39c}\ee
where $ \tilde\lambda^{M-}(z^+, u)$ is an unconstrained analytic
parameter. The remaining supervielbeins $H^{++M+},$ $ H^{++M-}$ and
$e^{--M}_\bA$
still transform according to  (\ref{26}),(\ref{27}) and (\ref{25}),
respectively, with parameters being the residual ones (\ref{39a}-\ref{39b}).

\section{The analytic frame solution}\label{solution}

We shall now show that in the `Frobenius' gauge (\ref{h-flat}) with
covariant derivatives taking the form
\begin{eqnarray} \D^+_\bA &=&  \p^+_{h\bA}  \cr
\D^-_\bA &=& - \p^-_{h\bA} + e^{--M}_\bA \p^+_{hM}  + \o^-_\bA \\[0pt]
\Dpp &=& \p^{++} + H^{++ M +}\p^-_{h M}
           + (x^{M +}_h + H^{++ M -}) \p^+_{hM}  + \o^{++}\
           ,\nonumber\label{h-deriv}\end{eqnarray}
the system of equations (\ref{28},\ref{29}), or equivalently the \ssdy
system (20) becomes manifestly soluble. In \cite{revisited} we showed that
for the $N=0$ case the \sd vierbein and connection are encoded in a
single unconstrained analytic prepotential, $\L^{+4}(x^+_h, u^\pm)$.
In the supersymmetric cases, this arbitrary datum gets generalised to
a charge $+4$ graded--skewsymmetric {\em supermultiplet} of analytic
{\em superfield} prepotentials $$\L^{+4NM} = \L^{+4[NM)}(z^+_h, u^\pm) =
\pmatrix{ \e^{\nu\mu} \L^{+4}(z^+_h, u^\pm) & \L^{+4\nu M}(z^+_h, u^\pm)\cr
- \L^{+4\nu M}(z^+_h, u^\pm)&\L^{+4(nm)}(z^+_h, u^\pm) }.$$
We shall now show that such
an unconstrained analytic prepotential, $\L^{+4NM} $,
encodes the general local solution of the self-dual supergravity
equations (20).

We begin with an arbitrary
{\it analytic} $H^{++\bB +}$ (satisfying (\ref{30})).
The relation (\ref{32}) then yields an expression for the harmonic
connection which is manifestly analytic, automatically satisfying its
equation of motion (\ref{31}),
\begin{equation}      \o^{++\bB}_\bA = \p^-_{h \bA} H^{++\bB +} .\label{40}\ee
Supertracelessness of this connection yields a local expression:
\begin{equation}      H^{++M+}  = (-1 )^{N} \p^-_{hN} \L^{+4MN}
  ,\label{41}\ee
yielding the required unconstrained analytic prepotential,
$\L^{+4MN} = \L^{+4[MN)}$.
The transformation rule (\ref{26}) induces the gauge invariance
$$\d \L^{+4MN} =
\dpp \l^{MN++} + H^{++[N+}\l^{M)+} + \l^{P+}\p^-_{h P} \L^{+4MN} +
 \p^-_{h P}\Lambda^{+5MNP} , $$
where $\l^{MN++}$ are the unconstrained gauge parameters in (\ref{39a})
and $\Lambda^{+4MNP}=\Lambda^{+4[MNP)}$ are parameters of pregauge
invariances of \r{41}. Unlike the $N=0$ case for which only the
first term on the right survives and we may choose a `normal gauge'
fixing this invariance \r1, for general $N$ it is not immediately clear how to
choose a representative example within this gauge-equivalence class of
prepotentials.

Eq.(\ref{33}) remains; from it we have
another expression for the harmonic connection
\begin{equation}   \o^{++\bB}_\bA = \p^+_{h \bA} H^{++ \bB -}\  .\label{42}\ee
Consistency of the two expressions (\ref{40},\ref{42}) for $ \o^{++\bB}_\bA $
yields a relationship between the two supervielbeins in $\Dpp$,
\begin{equation}     \p^+_{h \bA} H^{++ \bB -} = \p^-_{h \bA} H^{++\bB +},\ee
which may be solved for $H^{++ \bB -}$ in terms $ H^{++\bB +}$ thus:
\begin{equation}   H^{++ \bB -} = z_h^{-A} \p^-_{h \bA}  H^{++\bB +}
= (-1 )^C  z_h^{-A} \p^-_{h \bA} \p^-_{h \bC} \L^{+4BC} ,\label{43}\ee
up to an arbitrary analytic function, which can be set to zero using the
gauge freedom (\ref{27}).

We can therefore determine all the required fields ($H^{++M\pm}$
and $\o^{++}$) consistently, i.e. solve the dynamical content of (20), in
terms of the unconstrained (i.e. arbitrary) analytic prepotential
$\L^{+4NM}$.
As for the N=0 case \cite{revisited} , all the other
equations from (20) are also indeed solved in terms of $\L^{+4NM}$ and
determine the other analytic frame fields as functionals of $\L^{+4NM}$.
The proof follows that for the $N=0$ case given in \cite{revisited}.

\section{ The extraction  of central frame supervierbeins and
superconnections} \label{recipe}

As we have seen, the analytic prepotential ${\cal L}^{+4NM}$ encodes all the
analytic basis data. The procedure for  extracting the geometrical data
(supervierbein and superconnection) in the original central basis from
some specified analytic prepotential ${\cal L}^{+4NM}$ is as follows.

A. From (\ref{41}) and (\ref{43}) obtain the supervielbeins of $\D^{++}$:
$$\begin{array}{lcr} H^{++N+} &=& (-1 )^M  \p_{hM}^- {\cal L}^{+4NM} \cr
H^{++N-} &=& (-1 )^M z_h^{P -} \p^-_{h P} \p^-_{hM} {\cal L}^{+4NM} \ea $$

B. Consider (\ref{13}) as equations for the holomorphic coordinates
$z^{N +}_h$:
\be \p^{++} z^{N +}_h = (-1 )^M  \p^-_{hM} {\cal L}^{+4NM} .\ee
Integrating these first order equations find $z^{N +}_h$ as functions
of the {\it central} frame coordinates $z^{N \pm}
(\equiv z^{N a} u^\pm_a )$ and the harmonics.

C. Having obtained $z^{N +}_h$, similarly solve (\ref{14}),i.e.
\be \p^{++} z^{N -}_h =
  z^{N +}_h + (-1 )^M z_h^{\bA -} \p^-_{h \bA} \p^-_{hM} {\cal L}^{+4NM} \ee
in order to determine
$z^{N -}_h$ as a function of the central frame coordinates.

D. From (\ref{40}) obtain the connection of $\Dpp$:
\be \o^{++\bB}_\bA = (-1 )^M \p^-_{h \bA} \p^-_{hM} {\cal L}^{+4\bB M} ,\ee
and using the results of steps B and C, express it explicitly
in terms of central coordinates.

E. With the $\o^{++}$ obtained in step D, solve
\begin{equation}    \p^{++} \f = \f \o^{++}, \label{44}\ee
(i.e. equation (\ref{18}) rewritten in the central frame)
to obtain the bridge $\f$ in central coordinates.

F. Using results of steps B and C evaluate the supermatrices of
central coordinate partial derivatives $\dd{ z^{M \pm}}{z^{N -}_h}$
required for the transformation back to central coordinates.

\noindent  The above data affords the immediate construction of
explicit \sd supervielbeins  and connections as follows:

G. To transform the analytic frame $\D^+_M=\p^+_M$
back to the central frame, we clearly need to perform the
transformation
$\p^+_M \rightarrow \D^+_A = \f^\bA_A \dd{z^{M -}}{z^{\bA -}_h}\p^+_M$.
Therefore, multiply the bridge $\f$ obtained in step E with one of the
supermatrices from step F, and extract
the \sd supervielbein from the relation
\begin{equation}    Z \equiv  \f^\bA_A \dd{z^{M -}}{z^{\bA -}_h}
= u^{+a} E^{M b}_{A a}u^-_b\   ,\label{46}\ee
using the completeness relation
$ u^{+a} u_b^{-} - u^{-a} u_b^{+} = \d^a_b $.
The left-hand-side, as a function of the central frame
coordinates $\{ z^{M a}= z^{M+}u^{-a}-z^{M-}u^{+a} , u^\pm_a \}$,
is by construction bilinear in the harmonics, so
the supervierbeins $E^{M b}_{A a} =E^{M b}_{A a}(z^{N c})$ thus constructed
automatically depend only on the customary superspace coordinates (i.e. are
independent of the $u$'s).

H. The connection $\o_{A}^+$ is given in terms of the bridge by  the formula
(\ref{conn}), which therefore yields $\o_{A a} = \o_{A a}(z^{M b})$,
since $\o_{A}^+$, as a function of central frame coordinates
$\{ z^{M a} , u^\pm_a \}$, is by construction (see (\ref{c})) linear
in the harmonic $u^{+ a}$.

Therefore, extract the central frame \sd superconnection from the formula
\begin{eqnarray} (\o_{A}^+ ))^C_B &=& (\o_{A a})^C_B u^{+ a}  =
- D^+_A \f^\bB_B (\f^{-1})^C_\bB  \cr
&=& - \f^{\bC}_A D^+_{\bC} \f^\bB_B (\f^{-1})^C_\bB  =
 - \f^{\bC}_A \dd{ \f^\bB_B}{z^{\bC -}_h} (\f^{-1})^C_\bB \cr &=&
- \f^{\bC}_A ( \dd{z^{M +}}{z^{\bC -}_h}  \dd{\f^\bB_B }{z^{M +}}
              + \dd{z^{M -}}{z^{\bC -}_h}  \dd{\f^\bB_B }{z^{M -}} )
  (\f^{-1})^C_\bB  \;  .
\label{49}\end{eqnarray}
\section{An explicit example} \la{ex}

To conclude this paper, we illustrate the above procedure for a simple
monomial example of the analytic prepotential from which we explicitly
extract the \sd supervielbeins and superconnections. Further examples
will be given in a separate publication.
For the purely even case, the simplest example of prepotential is
$\L^{+4} \sim (x_h^{1+}x_h^{2+} )^2$ corresponding to the Euclidean Taub-NUT
space. By analogy  we shall consider the simple N=2 example
\be  {\cal L}^{+4MN} =
\pmatrix{g \e^{\mu\nu} \bt_h^{1+}\bt_h^{2+}x_h^{1+}x_h^{2+} & 0 \cr 0 & 0 },
\ee where $g$ is a parameter of dimension $[cm]^{-1}$. This yields
$$ \begin{array}{llcl}
 H^{++\mu+} &=& g \eta^{++}  (\s_3)_\nu^\mu  x_h^{\nu+} &= \dpp x_h^{\mu+}\;
 ;\qquad \eta^{++} \equiv  \bt_h^{1+}\bt_h^{2+}\;  ,\\[8pt]
H^{++i+}  &=& 0  &= \dpp \bt_h^{i+} .\ea $$
So $\bt_h^{i+} = \bt^{i+} = \bt^{ia} u^+_a $, the coordinates $\bt_h^{i+}$ are
flat, and
\be  x_h^{\mu+} =  (e^{g\eta \s_3} )_\nu^\mu  x^{\nu+}\; ;\qquad
\eta \equiv \half (\bt^{1-}\bt^{2+} + \bt^{1+}\bt^{2-}) .\ee
Similarly,
$$ \begin{array}{llcl}
 H^{++\mu-} &=& g (\s_3)_\nu^\mu (\eta^{++} x_h^{\nu-}  +  2\eta x_h^{\nu+})
 &= \dpp x_h^{\nu-} - x_h^{\nu+}\  ,\\[8pt]
 H^{++i-}  &=& 0  &= \dpp \bt_h^{i-} - \bt_h^{i+} , \ea $$
yield $\bt_h^{i-}= \bt^{i-} = \bt^{ia} u^-_a  $, so all the odd coordinates
are flat, and
\be  x_h^{\nu-} =
(e^{g\eta \s_3} )_\mu^\nu(x^{\mu-}+g(\s_3)_\mu^\nu x^{\mu+} \eta^{--})\  .\ee
where $\eta^{--} \equiv  \bt^{1-}\bt^{2-} .$
The inverse functions are then
\be \arr x^{\mu+} &=&  (e^{-g\eta \s_3})_\nu^\mu  x_h^{\nu+}\; \\[8pt]
        x^{\mu-} &=& (e^{-g\eta \s_3})_\nu^\mu
(x_h^{\nu-} - g(\s_3)_\nu^\mu  x_h^{\nu+} \eta^{--}) ,\ea\ee
since $\eta \eta^{--} = 0 $.
The central coordinate partial derivatives (step F) for the transformation
back to central coordinates are therefore given by
\def\x#1#2{(\s_3)_\rho^{#2} x_h^{\rho #1} }
$$  \dd{x^{\mu -}}{x_h^{\nu -}} =  (e^{-g\eta \s_3})_\nu^\mu  \;  \qquad
    \dd{\bt^{m -}}{\bt_h^{n-}} =  \d^m_n     \; \qquad
   \dd{x^{\mu -}}{\bt_h^{n -}} =\half g \x-{\mu} \bt^+_n + g \x+\mu \bt^-_n
    \;  ,$$
$$  \dd{x^{\mu +}}{\bt_h^{n-}} = \half g \x+{\mu} \bt^+_n \;  $$
with all other elements of the supermatrices
$  \dd{ z^{M \pm}}{z^{N -}_h} $ being zero. Since $N=2$ for this example,
we may lower the index of the odd coordinate using the $\e$ tensor thus:
$ \bt^\pm_i = \e_{ij} \bt^{j\pm} $.
The nonzero components of the superconnection $\o^{++}$ are
\begin{equation}    \begin{array}{lcl}
(\o^{++} )_\ba^\bb &=& \p^-_{h\ba} H^{++\bb+} =
                       g\eta^{++} (\s_3)_\ba^\bb \\[7pt]
(\o^{++} )_\bi^\bb &=& D^-_{\bi} H^{++\bb+} = - g \x+\bb \bt^+_\bi
		      \ea \la{opp} .\ee
Solving \r{44}, we obtain the bridge in central coordinates,
$$ \f_A^\bA = \pmatrix{ (e^{g\eta \s_3})_\a^\ba & 0 \cr
     - {g\over 2}(\s_3 e^{g\eta \s_3})_\mu^\ba x^{\mu (+}\bt^{-)}_i
                     &\d^\bi_i } ,$$
up to $\tau-$supergauge transformations \r{tau}.
The supervielbein, immediately extractable from the formula \r{46}, is
\begin{equation}    E^{M b}_{A a} = \pmatrix{ \d^\mu_\a \d^b_a & 0 \cr
      -{g\over 2} (\s_3)^\mu_\nu x^\nu_a \bt^b_j & \d^m_j \d^b_a }  \ee
and clearly describes a superspace with flat `body' and non-flat `soul'.
Evaluating the complex structures \r{cs}, we obtain the supermatrix
components
\begin{equation}   \arr (I^k )^{m a}_{n b} &=& - i\d^m_n (\s^k)^a_b\  ,\quad
        (I^k )^{\mu a}_{\nu b} =  - i\d^\mu_\nu (\s^k)^a_b\  ,\\[7pt]
    (I^k )^{m a}_{\nu b} &=& 0\  ,\qquad
    (I^k )^{\mu a}_{n b} =
      -i {g\over 2} (\s_3 )^\mu_\l ( x^\l_b \bt^c_n (\s^k)^a_c -
				    x^\l_c \bt^a_n (\s^k)^c_b )\  .\ea\ee
The non-zero superconnection components, extracted from \r{49} are
\begin{equation}   \arr (\o_{\a a})^\g_i
      &=& (\o_{ia})^\g_\a = - {g\over 2} (\s_3)^\g_\a \bt_{ia} \\[7pt]
	 (\o_{ia})^\g_j &=& {g\over 2}( (\s_3)^\g_\a x^\a_a \e_{ij} -
		         {g\over 2} x^{\g b} \bt_{ib}\bt_{ja})\;  .\ea\ee
{}From this supervielbein and superconnection, the vanishing of the
supertorsion
may be verified and the only nonzero components of the supercurvature tensor
\r{curv} may be found to be
\begin{equation}
 R_{\b ij}^{~~~~\a} = R_{ji\b}^{~~~~\a} = g (\s_3)^\a_\b \e_{ij}\  .
 \ee
Using \r{nc} the supercurvature components of the original non-chiral
theory may now be reconstructed:
$$ (W_{\a i} )_j^\b =  (Z_{ji} )^\b_\a = g (\s_3 )^\b_\a \e_{ij}\  ,\qquad
  (Z_{ij} )^\b_k   = 2 g (\s_3 )^\b_\a \e_{ij} \t^\a_k \      ,$$
all other components vanishing. We hope to return to a discussion of this
curious \hk superspace as well as further examples in the future.
\vskip 15 pt
We are very grateful to  D. Alekseevsky,  E. Ivanov
and D. Leites  for useful discussions
and the Erwin Schr\"odinger Institute, Vienna, where much of this work
was performed, for hospitality. One of us (V.O) also thanks the
Max-Planck-Institut f\"ur Mathematik, Bonn for hospitality during the
performance of the early stages of this work.

\end{document}